\documentclass[%
twocolumn,
footinbib,
amsmath,
amssymb,
aps,
pre
]{revtex4-2}

\usepackage{graphicx,xcolor}
\usepackage{dcolumn}
\usepackage{bm}
\usepackage{graphicx}
\usepackage{wrapfig}
\usepackage{float}
\usepackage{pgfplots}
\pgfplotsset{width=10cm,compat=1.9}

\newcommand{\treetilde}{\tilde}

\newcommand{\nocontentsline}[3]{}
\newcommand\stoptoc{%
   \let\origcontentsline\addcontentsline
   \let\addcontentsline\nocontentsline
}
\newcommand\resumetoc{%
   \let\addcontentsline\origcontentsline
}

\begin{document}

\preprint{APS/123-QED}

\title{Coherent modeling of double-folded ring polymers and their underlying random tree structure}

%
\author{Pieter H. W. van der Hoek}
\email{pvanderh@sissa.it}
\affiliation{
SISSA - Scuola Internazionale Superiore di Studi Avanzati, Via Bonomea 265, 34136 Trieste, Italy
}
\author{Elham Ghobadpour}
\email{elham.ghobadpour@ens-lyon.fr}
\affiliation{
ENS de Lyon, CNRS, Laboratoire de Physique (LPENSL UMR5672) et Centre Blaise Pascal, 69342 Lyon cedex 07, France
}
\author{Ralf Everaers}
\email{ralf.everaers@ens-lyon.fr}
\affiliation{
ENS de Lyon, CNRS, Laboratoire de Physique (LPENSL UMR5672) et Centre Blaise Pascal, 69342 Lyon cedex 07, France
}
\author{Angelo Rosa}
\email{anrosa@sissa.it}
\affiliation{
SISSA - Scuola Internazionale Superiore di Studi Avanzati, Via Bonomea 265, 34136 Trieste, Italy
}

\date{\today}

\begin{abstract}
Topologically constrained genome-like polymers often double-fold into tree-like configurations, which can be modelled on the level of folded (ring) polymers or on the level of the underlying random trees. 
For both descriptions, we have recently obtained expressions for the configurational entropy in ensembles with controlled branching activity.
Here we demonstrate that they are equivalent up to a contribution originating from the number of distinct wrappings of a single tree.
This allows us to develop a coherent framework for freely switching between the two representations.
Importantly, the equivalence extends to interacting systems provided the interactions are treated consistently on the tree and on the ring level.
To demonstrate the utility of the scheme, we introduce a generalization of the Amoeba Monte Carlo algorithm capable of generating the required ensembles of trees with fluctuating sizes.
While the tree algorithm reproduces results obtained by dynamic simulations of the corresponding ring model, it is ${\cal O}(N)$ faster for the purpose of sampling static properties and leverages the utility of the ring model for the study of dynamical properties, when used for the preparation of equilibrated starting states.
\end{abstract}

\maketitle

\stoptoc

\section{Introduction}\label{sec:Introduction}
\begin{figure}
\includegraphics[width=0.48\textwidth]{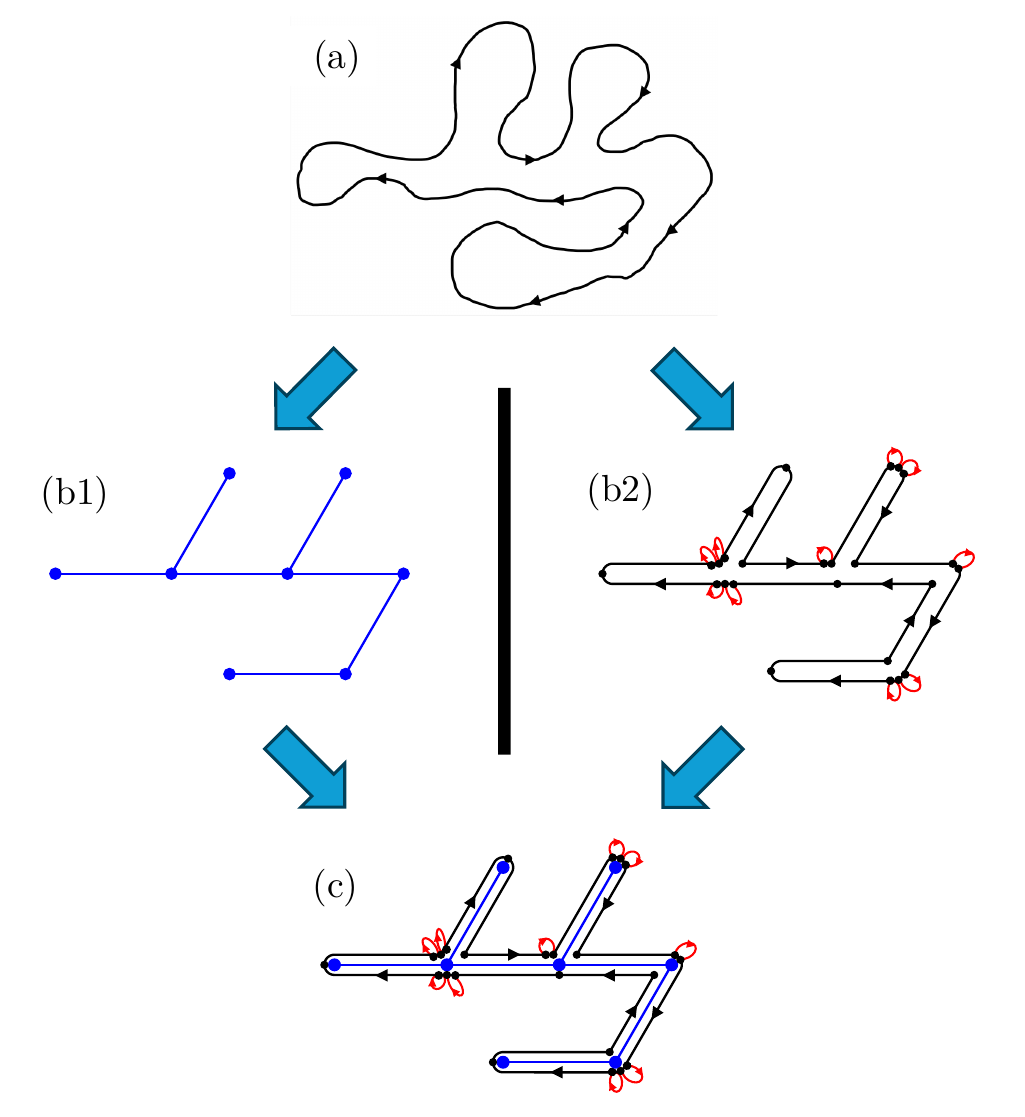}
\caption{
Illustration of a typical genome-like polymer adopting the conformation of a double-folded tree-like ring polymer (a).
We can choose to either model the underlying tree-structure (b1), or use elastic lattice models (b2) where two consecutive monomers of the ring can occupy the same spatial position of a tree monomer and the bond between them becomes a unit of stored length (red arcs).
The combined model (c) can either be inferred from wrapping a ring around the tree-structure (b1) or by inferring the tree-structure from the spatial positions of the ring-monomers (b2).
}
\label{fig:Modelfigure}
\end{figure}

Topologically constrained genome-like polymers often double-fold into tree-like configurations (Fig.~\ref{fig:Modelfigure}(a)) as
they form plectonemes due to supercoiling~\cite{MarkoSiggia1994,MarkoSiggiaSuperCoiledDNA1995,Woldringh1999,Cunha2001},
undergo loop extrusion~\cite{AlipourMarkoNAR2012,Sanborn2015,Fudenberg2016,GoloborodkoBJ2016,GoloborodkoELife2016},
or maximize the entropy of the crumpled~\cite{grosbergEPL1993} territorial~\cite{CremerReview2001} arrangement of interphase chromosomes arising from the decondensation of topologically untangled metaphase chromosomes~\cite{RosaPLOS2008,Dekker-Hic2009,RosaBJ2010}.
On the mesoscale, such systems can be described as randomly double-folding (ring) polymers~\cite{KhokhlovNechaev85,RubinsteinPRL1986,RubinsteinPRL1994,GrosbergSoftMatter2014,SmrekGrosberg2015}, a description that applies more widely in the context of dense solutions and melts of unknotted and non-concatenated ring polymers~\cite{SchroederRingsReview2025}.

In the past we have developed codes for simulating such systems either on the tree (Fig.~\ref{fig:Modelfigure}(b1)) or on the ring (Fig.~\ref{fig:Modelfigure}(b2)) level. 
In the first line of work, we have simulated (interacting) lattice trees using efficient Amoeba Monte Carlo (MC) algorithms that randomly cut and paste leaves from the trees~\cite{RosaEveraersPRL2014,Rosa2016a,Rosa2016b,Amoebapaper2024}. 
When focusing on the tree nodes as the relevant degrees of freedom, a wrapping procedure~\cite{RosaEveraersPRL2014,RosaEveraers2019} allows to generate corresponding ensembles of double-folded rings in a second step, which can be locally equilibrated when switching to a more microscopic fiber model~\cite{RosaEveraersPRL2014,SchramSM2019}.
In the second line of work~\cite{Ghobadpour2021,Ghobadpour2025}, we have studied an elastic lattice model of randomly branching, tightly double-folding ring polymers using a MC scheme where the dynamics of the monomers is local both in space and on the tree.
When focusing on the ring monomers as the relevant degrees of freedom, the corresponding trees can be either inferred {\it a posteriori}~\cite{Ghobadpour2021} or on the fly~\cite{Ghobadpour2025}.
The objective of the present work is the integration of the two approaches into a common, coherent framework (Fig.~\ref{fig:Modelfigure}(c)) identifying equivalent tree and ring ensembles and allowing us to switch at will from one representation and simulation method to the other.

The manuscript is structured as it follows:
In the Theory part (Sec.~\ref{sec:Theory}) we work out a bijective quantitative mapping between the representations on the ring and on the tree level, where we build on our recent exact results for the configurational entropy for the two descriptions in ensembles with controlled branching activity~\cite{vanderHoek2025,vanderHoekWrapping2026}.
In particular, 
(i) we demonstrate that the respective statistical weights in non-interacting systems are equivalent up to a contribution originating from the number of distinct wrappings of a single tree and
(ii) we argue that this relation continues to hold in general provided (volume) interactions are treated consistently across the two descriptions.
In the Methods Sec.~\ref{sec:Methods}, first we review briefly the MC lattice algorithm for ring polymer dynamics of Ref.~\cite{Ghobadpour2025} (Sec.~\ref{sec:ElhamsAlgorithm}), then we present (Sec.~\ref{sec:SamplingRings}) a generalization of the Amoeba MC algorithm~\cite{SeitzKlein1981,Rosa2016a,Rosa2016b,Amoebapaper2024} with decoupled cut and paste moves and capable of generating the required ensembles of trees of fluctuating sizes.
In the Results Sec.~\ref{sec:Results}, we compare the equilibration dynamics for the two algorithms for interacting rings, providing evidence that they generate indeed equivalent ensembles and compare their respective computational efficiency. 
We conclude with a brief summary and outlook in Sec.~\ref{sec:Conclusion}.

\section{Theory}\label{sec:Theory}

\subsection{Trees and wrapped rings}\label{sec:TreesDefinitions}
We consider double-folding of rings composed of length $N_{\rm ring}$ monomers around {\it acyclic} trees of $N_{\rm tree}$ nodes embedded on a common regular lattice of coordination number $c$ and unit step length $b$.
Specifically our implementations work on the $3d$ FCC lattice ($c=12$).

To be able to wrap a tree of $N_{\rm tree}$ nodes, a ring has to have a length $N_{\rm ring} \ge 2(N_{\rm tree}-1)$. 
In our elastic lattice model~\cite{Ghobadpour2021,Ghobadpour2025} rings are typically not extended to their maximal contour length and the
\begin{equation}\label{eq:Nrep}
N_{\rm rept} = N_{\rm ring} - 2(N_{\rm tree}-1)
\end{equation}
reptons (or zero-length bonds) represent freely displaceable units of stored lengths (cfr., the red arcs in panels (b2) and (c) of Fig.~\ref{fig:Modelfigure}).

As in past work~\cite{RosaEveraersPRL2014,Rosa2016a,Rosa2016b,Amoebapaper2024}, below we mostly focus on the case where the functionality $f$ of the tree nodes ({\it i.e.} the number of other nodes they are connected to) is limited to $f\le 3$ so that our trees are composed of $\{N_1, N_2, N_3\}$ nodes of functionality $f=1,2,3$, with 
\begin{eqnarray}
N_1(N_{\rm tree},N_3) & = & N_3+2 \, , \label{eq:n1_res} \\
N_2(N_{\rm tree},N_3) & = & N_{\rm tree} - 2N_3 - 2 \, . \label{eq:n2_res}
\end{eqnarray}
%

\subsection{Statistical ensembles of labelled trees with controlled branching activity}\label{sec:TotalNumberLabelledTrees}
In Ref.~\cite{vanderHoek2025} we have used Pr\"ufer counting to derive the multiplicity of configurations of trees with {\it independently labelled} nodes.
In the present context the identity of the tree nodes is defined through the ring monomers placed upon them.
Dividing out the irrelevant relabelling multiplicity of $N_{\rm tree}!$ the multiplicity of wrappable tree configurations as a function of $N_{\rm tree}$ and $N_3$ is given by~\footnote{Note that by dividing out $N_{\rm tree}!$, Eq.~\eqref{eq:labeled_trees} is not guaranteed to be an integer number.}
\begin{equation}\label{eq:labeled_trees}
\Omega_{\rm tree}(N_{\rm tree}, N_3) = \left\{
\begin{array}{cc}
1 \, , & \mbox{if} \,\,\, N_{\rm tree} = 1 \\
\\
\frac{(N_{\rm tree}-2)!}{N_1! \, N_2! \, N_3!} \frac1{2^{N_3}} \, , & \mbox{if} \,\,\, N_{\rm tree} \geq 2
\end{array}
\right.
\end{equation}
with $N_1$ and $N_2$ given by Eq.~\eqref{eq:n1_res} and Eq.~\eqref{eq:n2_res}, respectively.
Including the conformational entropy (neglecting interactions, each of the $N_{\rm tree}-1$ bonds specifying the spatial embedding can be oriented in one of $c$ ways) and controlling the number of branch-nodes via a chemical potential, $\mu_3$, the partition function reads 
\begin{equation}\label{eq:Z_tree}
Z_{\rm tree}(N_{\rm tree}, \mu_3) = c^{N_{\rm tree}-1} \sum_{N_3=0}^{N_{3, {\rm max}}} \Omega_{\rm tree}(N_{\rm tree},N_3) \, e^{\beta \mu_3 N_3} \, ,
\end{equation}
with $\beta^{-1} = k_BT$ where $T$ is the temperature and $k_B$ is the Boltzmann constant.

\subsection{Statistical ensembles of double-folding rings with controlled branching activity}\label{sec:NumberViableWrappingCodes}
In Ref.~\cite{vanderHoekWrapping2026} we have shown that the total number of tightly double-folded ring configurations wrapping a tree of given $(N_{\rm tree}, N_3)$ is
\begin{equation}\label{eq:labeled_tree_wrappings_labeled_ring}
\Omega_{\rm dfr}(N_{\rm tree}, N_3) =
\left\{
\begin{array}{cc}
1 \, , & \mbox{if} \,\,\, N_{\rm tree} = 1 \\
\\
\frac{2 \, (N_{\rm tree} -1)!}{N_1! \, N_2! \, N_3!} \, , & \mbox{if} \,\,\, N_{\rm tree} \geq 2 \\
\end{array}
\right.
\end{equation}
with, again, $N_1$ and $N_2$ given by Eq.~\eqref{eq:n1_res} and Eq.~\eqref{eq:n2_res}, respectively.
In addition, the reptons in our elastic lattice model~\cite{Ghobadpour2021,Ghobadpour2025} can be placed in
\begin{equation}\label{eq:ElasticModel-ChoosingBonds}
\Omega_{\rm rep}(N_{\rm ring},N_{\rm tree})  = \frac{N_{\rm ring}!}{(2(N_{\rm tree}-1))! \, N_{\rm rept}!}
\end{equation}
distinct ways between the extended bonds linking ring monomers located on adjacent tree nodes.
Including the conformational entropy of the extended bonds and controlling again the number of branch-nodes via a chemical potential, $\treetilde\mu_3$, the partition function reads 
\begin{align}\label{eq:Z_ring}
& Z_{\rm ring}(N_{\rm tree}, N_{\rm ring}, \treetilde\mu_3) = c^{N_{\rm tree}-1} \times \nonumber\\
& \sum_{N_3=0}^{N_{3, {\rm max}}} \Omega_{\rm dfr}(N_{\rm tree},N_3) \, \Omega_{\rm rep}(N_{\rm ring},N_{\rm tree}) \, e^{\beta \treetilde\mu_3 N_3} \, .
\end{align}
%

\subsection{The wrapping entropy}\label{sec:WrappingEntropy}
Following the procedure originally outlined in~\cite{RosaEveraers2019}, a tree can be wrapped by a double-folded ring in
\begin{equation}\label{eq:wrapping multiplicity}
\Omega_{\rm wrap}(N_{\rm tree}, N_3) = \left\{
\begin{array}{cc}
1 \, , & \mbox{if} \,\,\, N_{\rm tree} = 1 \\
\\
2 (N_{\rm tree}-1) \times 2^{N_3} \, , & \mbox{if} \,\,\, N_{\rm tree} \geq 2 \\
\end{array}
\right.
\end{equation}
distinct ways, where the factor $2 (N_{\rm tree}-1)$ denotes the number of circular permutations of the ring ({\it i.e.} the possibility of choosing which ring monomer gets the label $i=1$) and the factor $2^{N_3}$ enumerates the different directions of wrapping that can be chosen when encountering a branch-node. 
Inspection of Eqs.~\eqref{eq:labeled_trees} and~\eqref{eq:labeled_tree_wrappings_labeled_ring} shows that
\begin{equation}\label{eq:multiplicity relation}
\Omega_{\rm dfr}(N_{\rm tree}, N_3) = \Omega_{\rm tree}(N_{\rm tree}, N_3) \, \Omega_{\rm wrap}(N_{\rm tree}, N_3) \, .
\end{equation}
Reassuringly the counting of Pr\"ufer codes of labelled trees~\cite{vanderHoek2025} and of wrapping codes for double-folding rings~\cite{vanderHoekWrapping2026} lead to equivalent results.

\subsection{Quantitative link between tree and ring ensembles}\label{sec:TreeRingSwitching}
We are now in the position to establish exact correspondences between descriptions on the tree and on the ring level for non-interacting systems.

One coherent choice is to establish a correspondence between {\it one} tree conformation and {\it an ensemble} of $\Omega_{\rm rep}(N_{\rm ring}, N_{\rm tree}) \, \Omega_{\rm wrap}(N_{\rm tree}, N_3)$ ring conformations.
Then, according to Eqs.~(\ref{eq:Z_tree}), (\ref{eq:Z_ring}) and (\ref{eq:multiplicity relation}), one must choose the two branching chemical potentials $\mu_3$ and $\treetilde\mu_3$ equal to each other.

Below we will adopt a different choice, where {\it one} tree conformation corresponds to {\it one} ring conformation.
This is possible, if we include the additional statistical weight
\begin{align}\label{eq:tree weight}
& w_{\rm dfr}(N_{\rm tree}, N_{\rm ring}, N_3) \nonumber\\
& =  \Omega_{\rm rep}(N_{\rm ring},N_{\rm tree}) \, \Omega_{\rm wrap}(N_{\rm tree}, N_3)
\end{align}
into our MC simulations of trees: tree conformations corresponding to a larger multiplicity of ring wrappings and repton distributions are generated with a correspondingly increased probability. 
With Eq.~\eqref{eq:tree weight}, including the branching weight and the weight for tree size $w_{\rm tree}(N_{\rm tree}, N_{\rm ring})$,
\begin{align}\label{eq:tree weight 2}
& w_{\rm tree}(N_{\rm tree}, N_{\rm ring}) \, e^{\beta \mu_3 N_3} \equiv w_{\rm dfr}(N_{\rm tree}, N_{\rm ring}, N_3) \, e^{\beta \treetilde\mu_3 N_3} \nonumber\\
& = \Omega_{\rm rep}(N_{\rm ring},N_{\rm tree}) \, \Omega_{\rm wrap}(N_{\rm tree}, N_3) \, e^{\beta \treetilde\mu_3 N_3} \nonumber\\
& = \frac{N_{\rm ring}! \, (2(N_{\rm tree}-1))}{(2(N_{\rm tree}-1))! \, (N_{\rm ring}-2N_{\rm tree} +2)!} \, 2^{N_3} \, e^{\beta \treetilde\mu_3 N_3} \nonumber\\
& = \frac{N_{\rm ring}!}{(2N_{\rm tree}-3)! \, (N_{\rm ring}-2N_{\rm tree} +2 )!} \, e^{(\beta \treetilde\mu_3 + \ln(2))N_3} \, , \nonumber\\
\end{align}
this implies the choices:
\begin{align}
& w_{\rm tree}(N_{\rm tree}, N_{\rm ring}) \nonumber\\
& =
\left\{
\begin{array}{cc}
1 \, , & \mbox{if} \,\,\, N_{\rm tree} = 1 \\
\\
\frac{N_{\rm ring}!}{(2N_{\rm tree}-3)! \, (N_{\rm ring}-2N_{\rm tree} +2 )!} \, , & \mbox{if} \,\,\, N_{\rm tree} \geq 2
\end{array}
\right.
\label{eq:w_tree}
\end{align}
and
\begin{equation}\label{eq:muring_mutree}
\beta\mu_3 = \beta\treetilde\mu_3 + \ln(2) \, .
\end{equation}
Eq.~\eqref{eq:muring_mutree} prescribes a simple shift between the chemical potentials to be used with the two representations as a convenient way to account for the ``entropy of wrapping'' (Sec.~\ref{sec:WrappingEntropy}) of double-folded rings with respect to trees.
The statistical weight~\eqref{eq:w_tree} controls the average tree size $\langle N_{\rm tree} \rangle$ through the competition between the tree and the repton entropies~\cite{vanderHoekWrapping2026}.
While the former increases with $N_{\rm tree}$, the latter is maximal for $N_{\rm rept} \simeq 2N_{\rm tree} \simeq N_{\rm ring}/2$ at fixed $N_{\rm ring}$ and then decreases (see Eqs.~\eqref{eq:Nrep} and~\eqref{eq:ElasticModel-ChoosingBonds}) due either to the corresponding reduction in the number of reptons for $N_{\rm rept} \lesssim N_{\rm ring}/2$ or to the reduction of the number of tree nodes on where to place them for $N_{\rm tree} \lesssim N_{\rm ring}/4$. 

\begin{figure*}
\includegraphics[width=1.00\textwidth]{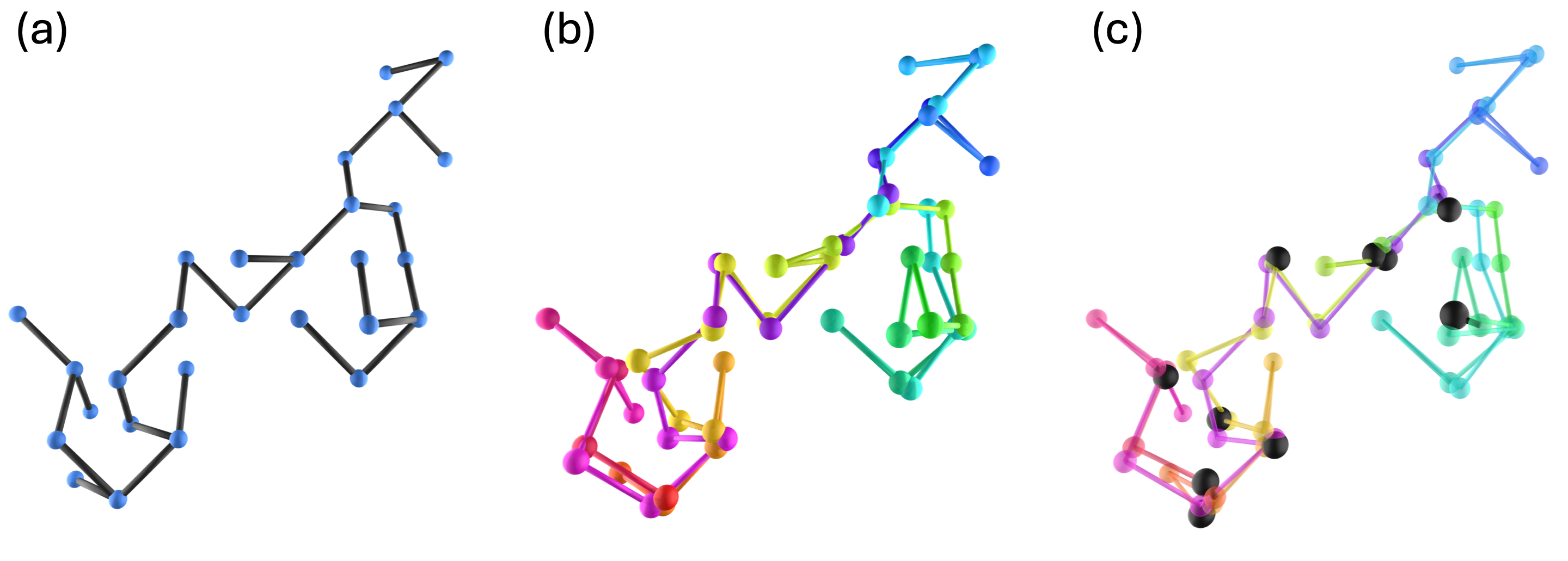}
\caption{
Visualization of the correspondence between the tree and ring representations for an isolated self-avoiding double-folded ring polymer with $N_{\rm ring} = 64$ on the fcc lattice in $d=3$ dimensions.
(a) Underlying tree structure.
(b) Wrapped ring representation, with cyclic coloring along the ring contour to visualize the continuous polymer path on the tree backbone.
(c) Full wrapped ring including reptons associated with zero-length bonds (shown in black) in the elastic lattice model.
In the two ring representations of panels (b) and (c), a small random positional displacement was applied to the monomer coordinates to improve the visibility of the local double-folded structure and overlapping segments
}
\label{fig:ExampleRing}
\end{figure*}

Importantly, adopting Eqs.~\eqref{eq:w_tree} and~\eqref{eq:muring_mutree}, we can freely switch between the use of the ring and of the tree representation.
In going from the ring to the tree description the information on repton locations and the sense of wrapping is simply eliminated.
To transform a tree into a double-folded ring, repton locations and the directions of wrapping at branch points are chosen randomly (Fig.~\ref{fig:ExampleRing}).

\subsection{Inclusion of excluded volume interactions}\label{sec:IncludingInteractions}
The correspondence between the descriptions on the tree and on the ring level continues to hold in the presence of interactions, provided {\it all} other contributions to the ring and tree Hamiltonians are calculated identically on the {\it tree} level. 
Below we will demonstrate this for the example of hard excluded volume interactions, {\it i.e.} imposing that two tree nodes cannot occupy the same lattice position.
Mathematically, this is defined introducing an interaction term
\begin{equation}\label{eq: hard excl vol}
\beta u_{\rm EV}(\vec r_i, \vec r_j) =
\left\{
\begin{array}{cc}
+\infty \, , & \mbox{if } \vec r_i = \vec r_j \\
\\
0 \, , & \mbox{otherwise}
\end{array}
\right.
\end{equation}
and corresponding statistical weight $w_{\rm EV} = \exp(-\beta u_{\rm EV})$ or
\begin{equation}\label{eq: hard excl vol weight}
w_{\rm EV}(\vec r_i, \vec r_j) =
\left\{
\begin{array}{cc}
0 \, , & \mbox{if } \vec r_i = \vec r_j \\
\\
1 \, , & \mbox{otherwise}
\end{array}
\right.
\end{equation}
where $\vec r_i$ and $\vec r_j$ are the lattice spatial coordinates of any pair of tree nodes $i$ and $j$ such that
\begin{equation}
w_{\rm EV} = e^{-\beta u_{\rm EV}} \, ,
\end{equation}
with
\begin{equation}
w_{\rm EV} = \prod_{i<j} w_{\rm EV}(\vec r_i, \vec r_j)
\end{equation}
and 
\begin{equation}
\beta u_{\rm EV} = \sum_{i<j} \beta u_{\rm EV}(\vec r_i, \vec r_j) \, .
\end{equation}

\section{Monte Carlo methods}\label{sec:Methods}
We have implemented various sampling schemes for double-folded rings and corresponding trees on the fcc lattice in $d=3$ dimensions.
Visualizations of a self-avoiding tree and corresponding double-folded rings with and without reptons are shown in Fig.~\ref{fig:ExampleRing}.
As stated above, in going from the ring to the tree description (Fig.~\ref{fig:ExampleRing} (c)$\rightarrow$(a)) we simply eliminate the information on repton locations and the order in which the tree branches are wrapped.
To transform a tree into a double-folded ring~\cite{RosaEveraersPRL2014,RosaEveraers2019} (Fig.~\ref{fig:ExampleRing} (a)$\rightarrow$(c)), repton locations and the directions of wrapping at branch points are chosen randomly.
While we have not implemented this, it is it worth pointing out that simply going back and forth between the tree and the ring levels corresponds in itself to a powerful non-local MC move for ring based simulations.

In Ref.~\cite{vanderHoek2025} we have shown how to efficiently generate non-interacting trees with controlled branching activity by randomly sampling Pr\"ufer codes from the counting scheme underlying Eq.~\eqref{eq:labeled_trees}.
Sampling a tree of $N_{\rm tree} = 10^6$ nodes takes of the order of one CPU second.
In principle, one could devise an analogous algorithm for generating randomly double-folded rings based on sampling wrapping codes from the counting scheme underlying Eq.~\eqref{eq:wrapping multiplicity}.
A simple sampling scheme would be less efficient, since only a fraction of $\mathcal{O}(1/N_3)$ of randomly generated wrapping codes is valid as a result of the ``ballot theorem'' attrition due to premature ring closure~\cite{vanderHoekWrapping2026}. 
While this could probably be overcome by (advanced) Rosenbluth methods~\cite{Rosenbluth1955,HsuGrassbergerReview2011}, we have not tried to implement such an algorithm, because we see not advantage compared to randomly wrapping Pr\"ufer sampled trees.

Below we focus on two complementary Metropolis~\cite{Metropolis1953} importance sampling algorithms for double-folded rings and trees which remain valid in the presence of (excluded) volume interactions. In particular, we highlight the changes we had to make compared to some of our earlier works~\cite{RosaEveraersPRL2014,SchramSM2019,Ghobadpour2021} to fully implement the present coherent framework for the two level description:
\begin{enumerate}
\item
When simulating (the dynamics of) double-folded rings with the ring monomers as degrees of freedom, we have to simultaneously keep track of the evolution of the ring and of the tree conformation as soon as the branching and/or volume interactions are defined on the tree level.
For the sake of completeness we review in Sec.~\ref{sec:ElhamsAlgorithm} how we have implemented this in Ref.~\cite{Ghobadpour2025}.
\item
When simulating lattice trees representing double-folded rings of fixed lengths $N_{\rm ring}$, we require an algorithm which allows the number of tree nodes, $N_{\rm tree}$, to fluctuate such as to reproduce the variations in the number of reptons in the elastic lattice model~\cite{Ghobadpour2025} (Sec.~\ref{sec:TreesDefinitions}). 
This is not possible with the available leaf-mover Amoeba algorithms~\cite{SeitzKlein1981,RosaEveraersPRL2014,Rosa2016a,RosaEveraers2019,Amoebapaper2024}, which conserve tree weights, because leaves (namely, endpoints or tree nodes with ``$f=1$'') are first cut and then pasted within each MC move.
To remove this constraint, we propose in Sec.~\ref{sec:SamplingRings} and in Sec.~\ref{sec:SemiGCAmoeba} in Supplemental Material (SM)~\cite{SMnote} a semi-grand canonical version of the Amoeba algorithm which decouples the cutting and pasting of leaves.
\end{enumerate}
%

\subsection{Simulating the local dynamics of randomly double-folded rings}\label{sec:ElhamsAlgorithm}
For studying the dynamics on the ring level we use the dynamic MC code introduced in Ref.~\cite{Ghobadpour2025} for the elastic lattice model of double-folded randomly branching ring polymers.
Trial moves consist in random displacements of randomly chosen monomers to adjacent lattice sites. 
Most of the moves are rejected for disrupting the double-folded structure characterising our rings. 
The two admissible moves are:
\begin{description}
\item[Repton moves] 
where monomers hop longitudinally along the ring, crossing an extended bond to a neighboring tree node, thereby transporting a unit of stored length along the tree without changing its structure or the system's energy.
\item[Hairpin moves]
which create, extend or remove side branches and modify $N_{\rm tree}$.
The trees grow when monomers flanked by two reptons hop in a transverse direction, thereby unfolding both zero-length bonds and generating a new pair of oppositely oriented extended bonds.
As a result the moved monomer is reassigned to a new tree node of functionality $f=1$, which is connected to the node to which the monomer was originally associated.
The latter's functionality and the total number of tree nodes, $N_{\rm tree}$, increase by one.
The corresponding inverse move removes an extended bond pair, leading to the shortening or complete removal (annihilation) of a side branch.
\end{description}
To keep track of both, ring and tree conformation, the data structure comprises:
(i) the spatial coordinates of ring monomers and tree nodes;
(ii) the information about which monomers belong to which tree node;
(iii) (implicit) information on the fixed ring connectivity and the dynamically changing connectivity of the graph describing the tree.
By convention, all monomers belonging to the same tree node occupy the same position in space, while neighboring nodes on the tree occupy adjacent lattice sites and are connected by a pair of oppositely oriented extended bonds.

\subsection{Sampling rings from the semi-grand canonical ensemble of the underlying tree conformations}\label{sec:SamplingRings}
On the tree level, the conformations generated by the above dynamic algorithm correspond to a ``semi-grand canonical'' ensemble of a fixed number of trees of fluctuating sizes, while standard Amoeba algorithms~\cite{SeitzKlein1981,RosaEveraersPRL2014,Rosa2016a,RosaEveraers2019,Amoebapaper2024} generate ensembles of a fixed number of trees of fixed size.
In Sec.~\ref{sec:SemiGCAmoeba} in SM~\cite{SMnote} we describe a variant which decouples the cutting and pasting of leaves.

The algorithm is largely based on the ``semi-kinetic'' version of the Amoeba algorithm introduced by us in Ref.~\cite{Amoebapaper2024} which allows for efficient trial move generation in the low- and moderate-branching regime of $\mu_3 < 0$, owing to a careful choice of transition probabilities satisfying the condition of detailed balance (see Eq.~\eqref{acc1} in SM~\cite{SMnote}).
Specifically the algorithm proposes two distinct moves, (a) one for leaf-cutting and (b) one for leaf-pasting to a ``$f=1$''-node or a ``$f=2$''-node in order to preserve the condition of $f\leq 3$ for nodes' functionality (see Sec.~\ref{sec:TreesDefinitions}).
The last move requires that the spatial position of the pasted node is chosen randomly amongst the $c$ nearest neighbors of the other node, with the additional prescription for excluded volume (see Sec.~\ref{sec:IncludingInteractions}) that two tree nodes cannot occupy the same lattice site.
Full details of the algorithm, which may be skipped on first reading, are provided in Sec.~\ref{sec:SemiGCAmoeba} in SM~\cite{SMnote}.

As pointed out in the introduction of this Section, in the final step before the further analysis, trees generated with the semi-grand canonical Amoeba algorithm (Fig.~\ref{fig:ExampleRing}(a)) are first randomly wrapped~\cite{RosaEveraersPRL2014,RosaEveraers2019} by a ring of minimal length (Fig.~\ref{fig:ExampleRing}(b)) and subsequently the required numbers of reptons are placed stochastically (Fig.~\ref{fig:ExampleRing}(c)).

\section{Results}\label{sec:Results}

\begin{figure*}
\includegraphics[width=0.95\textwidth]{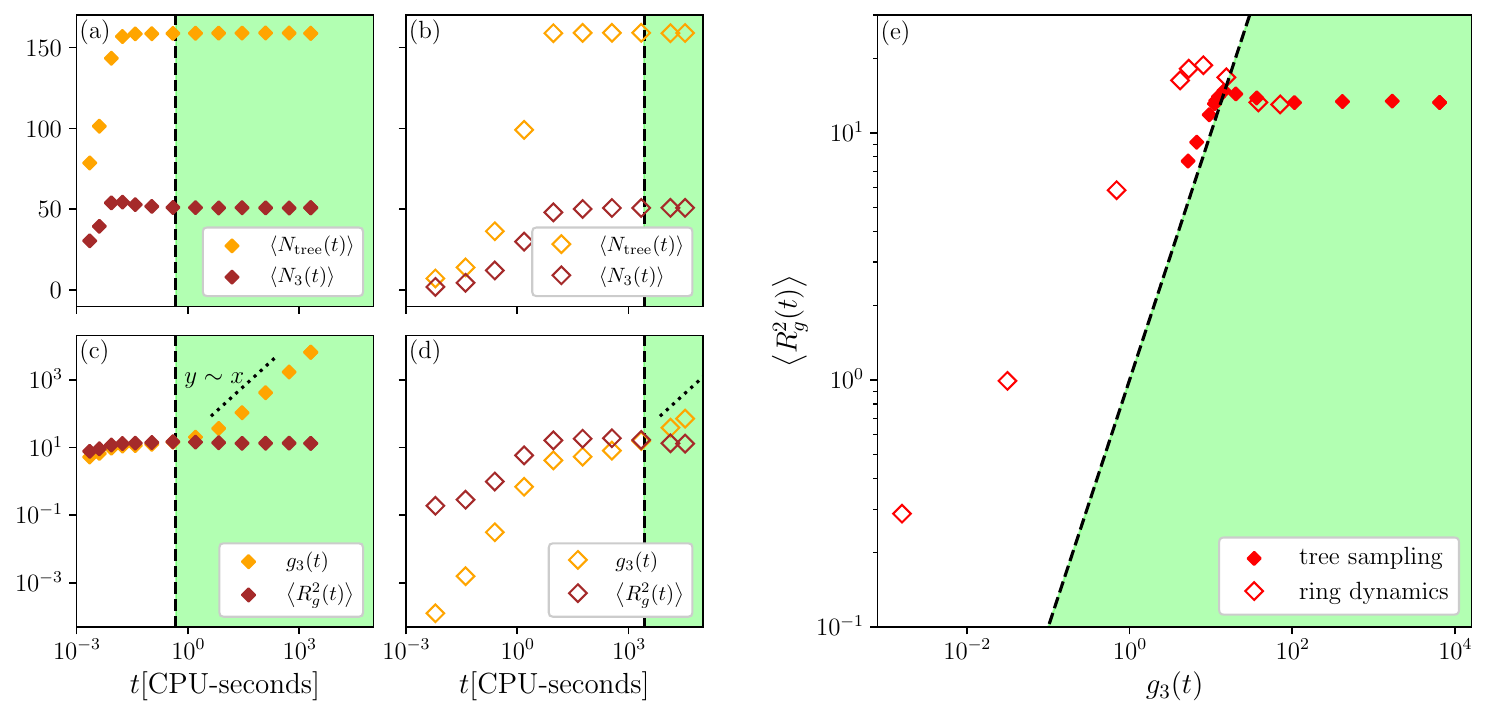}
\caption{
Comparing polymer equilibration for ring dynamics simulations (Sec.~\ref{sec:ElhamsAlgorithm}, open symbols) and Amoeba tree sampling (Sec.~\ref{sec:SamplingRings}, filled symbols).
Data here are for melts of tightly double-folded rings with $N_{\rm ring} = 512$ and $\beta \treetilde\mu_3 = 0$.
(a,b)
Mean number of tree nodes, $\langle N_{\rm tree}(t)\rangle$, and mean number of ``$f=3$''-nodes, $\langle N_3(t)\rangle$, as a function of time ($t$, in CPU-seconds).
(c,d)
Mean-square gyration radius, $\langle R_g^2(t) \rangle$ (Eq.~\eqref{eq:<Rg2>}), and mean-square displacement of the chain center of mass, $g_3(t)$ (Eq.~\eqref{eq:g3}), as a function of $t$.
Where $g_3(t) \gtrsim \langle R_g^2(t) \rangle$ (on the right-hand-side of the vertical dashed line, green-shaded area) polymers are equilibrated and move diffusively (see dotted line in panel (c)).
(e)
In the representation $(g_3(t)=x, \langle R_g^2 \rangle=y)$, polymer equilibration corresponds to the region $y\leq x$ (green-shaded area).
}
\label{fig:Rg_G3_explained}
\end{figure*}

\begin{figure*}
\includegraphics[width=1.00\textwidth]{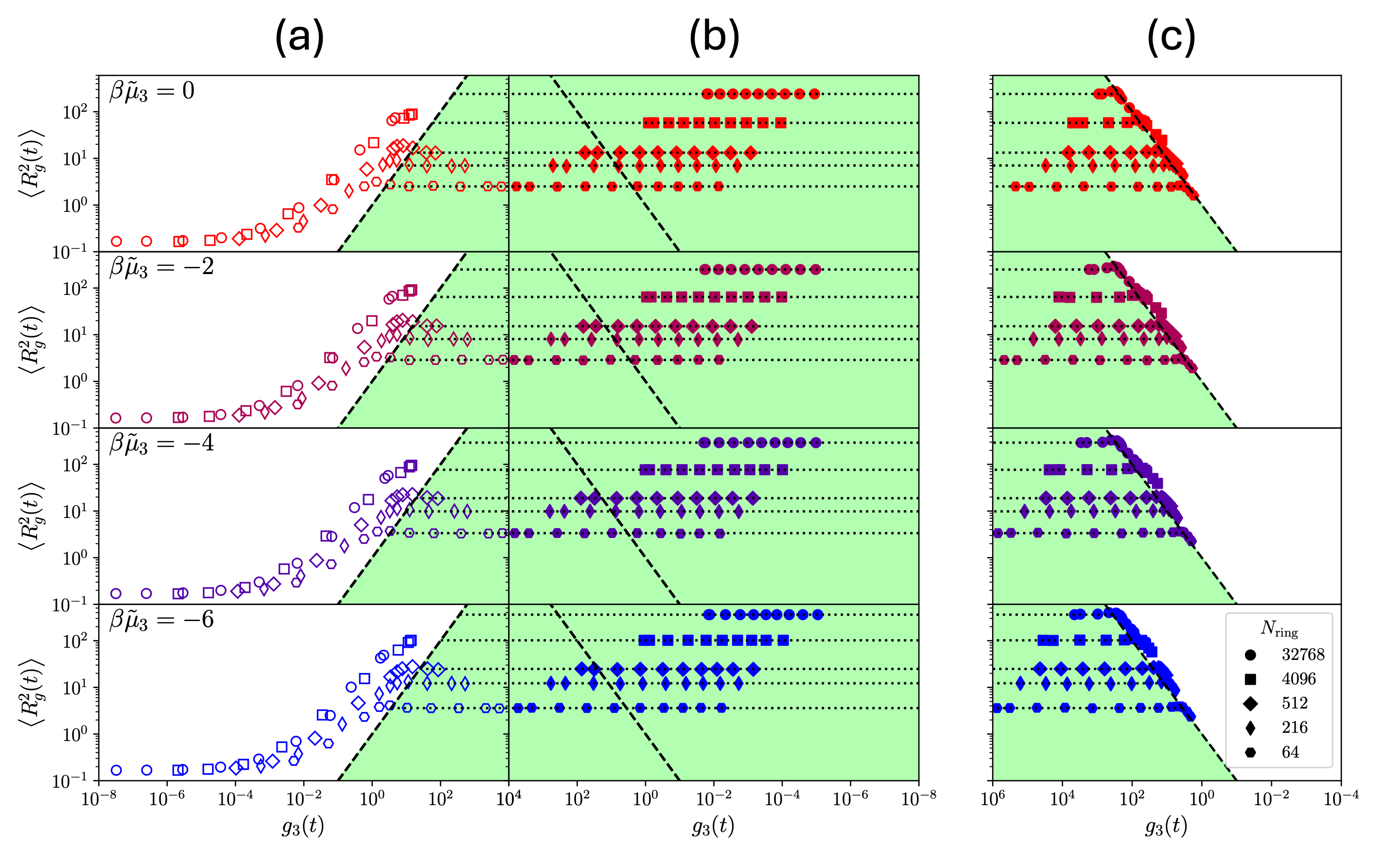}
\caption{
Time evolution of ring mean-square gyration radius ($\langle R_g^2(t) \rangle$, Eq.~\eqref{eq:<Rg2>}) {\it vs.} mean-square displacement of the ring centre of mass ($g_3(t)$, Eq.~\eqref{eq:g3}).
(a,b)
Ring dynamics simulations, with the following initial conformations:
(a) compact dimers (open symbols, time increases from left to right),
(b) equilibrated rings generated using the Amoeba tree sampling (filled symbols, time increases from right to left).
(c)
Amoeba tree sampling, with compact dimers for initial conformations.
Symbols of different colors are for different $\beta \treetilde\mu_3$ (values are in the figure), while different symbols are for different $N_{\rm ring}$ (see legend in the figure).
Dashed lines correspond to $y=x$: crossing this line marks the regime $g_3(t) \gtrsim \langle R_g^2(t) \rangle$ where chain conformations reach equilibrium (see Fig.~\ref{fig:Rg_G3_explained}(e)). 
The portions of the trajectories corresponding to equilibrated chains are green-shaded for clarity; notice the stationary behavior of $\left \langle R_g^2(t) \right\rangle$ for the chains prepared with the Amoeba protocol. 
Note: because both algorithms only start writing data after one MC ``sweep" (here, one MC sweep corresponds to the number of MC moves equal to the total number of monomers of the system), data for very short time scales are absent in Amoeba simulations.
}
\label{fig:Rg_G3}
\end{figure*}

\begin{figure*}
\includegraphics[width=0.98\textwidth]{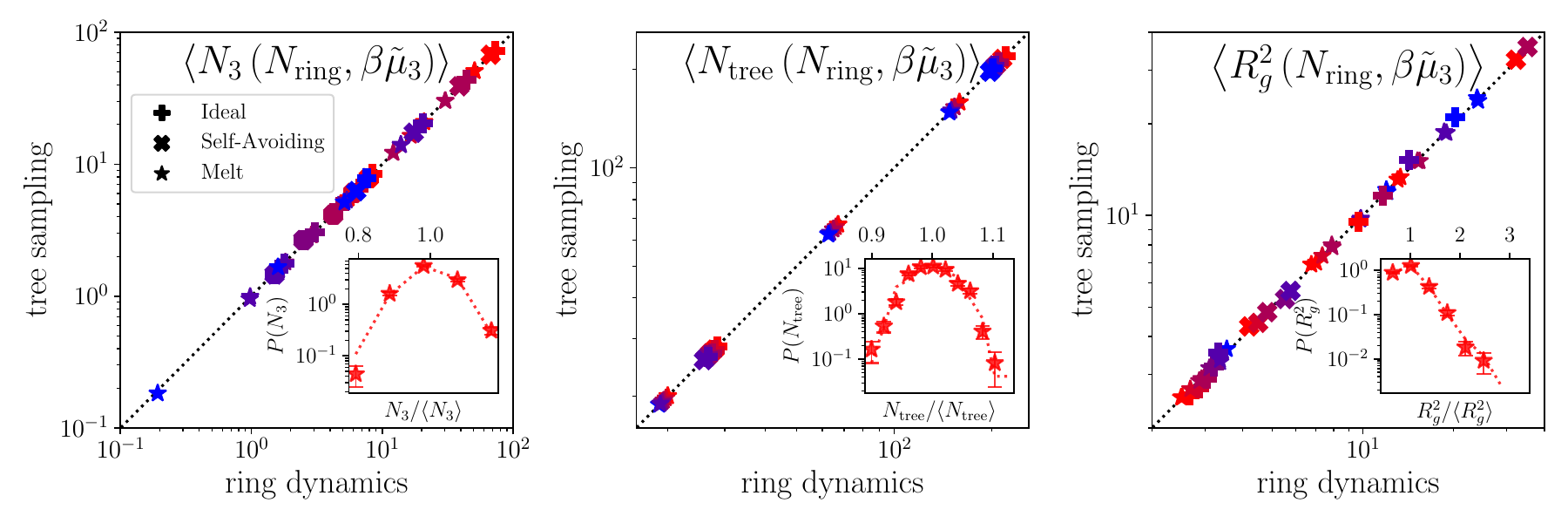}
\caption{
Comparing the properties of equilibrated ring polymers obtained by ring dynamics ($x$-axis) and Amoeba tree sampling ($y$-axis).
Results for: mean number of branch-points ($\langle N_3 \rangle$, left panel), mean tree-size ($\langle N_{\mathrm{tree}} \rangle$, middle panel) and mean-square gyration radius ($\langle R_g^2 \rangle$, right panel). 
Datapoints correspond to different values of input parameters ($N_{\rm ring}$, $\beta\treetilde\mu_3$), and excluded-volume conditions (ideal, dilute self-avoiding and melt, see legend). 
Colorcode is as in Fig.~\ref{fig:Rg_G3} ({\it i.e.}, ``red$\to$blue'' means ``more$\to$less'' branched trees; moreover, with respect to Fig.~\ref{fig:Rg_G3} we include here additional data for $\beta\treetilde\mu_3=-3,-1$).
All data collapse onto the $y=x$ line, confirming the equivalence of the two sampling methods.
The insets show corresponding probability distribution functions for ring melts with $N_{\rm ring}=512$ and $\beta \tilde{\mu}_3=0$, for ring dynamics (dotted lines) and Amoeba sampling (symbols). 
Distributions for other values of $N_{\rm ring}$ and $\beta \tilde{\mu}_3$ (not displayed) show equally excellent agreement.
}
\label{fig:Diagonal_plots}
\end{figure*}

\subsection{Systems}\label{sec:Systems}
We have generate ensembles of 
(i) ideal tightly double-folded rings in the absence of excluded volume interactions,
(ii) isolated self-avoiding tightly double-folded rings
and
(iii) melts of tightly double-folded rings
using both, the algorithm for simulating the ring dynamics (Sec.~\ref{sec:ElhamsAlgorithm}) and the Amoeba algorithm for sampling the corresponding tree ensemble (Sec.~\ref{sec:SamplingRings}).
For the three classes of systems, we have explored ring size values $N_{\rm ring} = \{ 64, 216, 512, 4096, 32768 \}$ for branching chemical potentials $\beta\treetilde\mu_3$ in the range $[-6, 0]$ using two different initial conditions: a compact starting state and an equilibrated ensemble.
The compact initial state consisted of tightly double-folded rings whose underlying tree structure were just two nodes located on neighboring lattice sites ({\it i.e.} dimers).
The equilibrated starting states were obtained from wrapped trees generated using Amoeba tree sampling.
For a detailed overview systems' sizes and simulation parameters, we refer the reader to the tables in Sec.~\ref{sec:input} in SM~\cite{SMnote}.

\subsection{Observables}\label{sec:Observables}
To validate our scheme and to estimate the computational efficiency of the two algorithms we observe the temporal evolution of ensemble averages of
the total number of tree nodes, $\langle N_\mathrm{tree}(t) \rangle$,
the number of branch nodes, $\langle N_3(t) \rangle$, 
the ring mean-square gyration radius,
\begin{equation}\label{eq:<Rg2>}
\langle R_g^2(t) \rangle \equiv \left\langle \frac1{N_{\rm ring}} \sum_{i=1}^{N_{\rm ring}} (\vec r_i(t) - \vec r_{\rm cm}(t))^2 \right\rangle \, ,
\end{equation}
where $r_{\rm cm}(t) = \frac1{N_{\rm ring}} \sum_{i=1}^{N_{\rm ring}} \vec r_i(t)$ is the ring center of mass position at time $t$, 
as well as the mean-square displacement of the ring centres of mass (CM),
\begin{equation}\label{eq:g3}
g_3(t) \equiv \langle ( {\vec r}_{\rm cm}(t) - {\vec r}_{\rm cm}(0) )^2 \rangle \, .
\end{equation}
%

\subsection{Equilibration}\label{sec:equilibration}
To set the stage, consider the example of a melt of interacting rings with $N_{\rm ring} = 512$ and $\beta \treetilde\mu_3 =0$.
In Fig.~\ref{fig:Rg_G3_explained} we show the temporal evolution of our observables for simulations starting from extremely compact dimer configurations, namely far from equilibrium.
There are several points to note:
(i) structural observables equilibrate ({\it i.e.}, reach a time-independent stationary value) on the time scale where the CM diffusion reaches the asymptotic diffusive regime, $g_3(t) \sim t$ (panels (c,d));
(ii) they equilibrate to equal values for ring and tree simulations (open and filled symbols, respectively);
(iii) the relaxation to the equilibrium value is not necessarily monotonic; 
(iv) equilibration is orders of magnitude faster for Amoeba tree sampling than for ring dynamics simulations.

In particular, equilibrium is reached when the CMs have diffused over distances exceeding the (asymptotic) ring or tree radius of gyration, $g_3(t) \gtrsim \langle R_g^2(t) \rangle$.
Thus while the (equilibration or decorrelation) time needed to decide whether a simulation is long enough is not known {\it a priori}, there is a built-in measure that allows to judge if our trees or rings have diffused over a sufficiently large distance: the mean-square gyration radius.
As in the past~\cite{Rosa2016a,Rosa2016b,Ghobadpour2021,Amoebapaper2024,Ghobadpour2025}, we thus prefer a representation of $\langle R_g^2(t) \rangle$ plotted against $g_3(t)$, where reaching a plateau in the green-shaded area beyond the ``$y=x$''-line is indicative of true equilibration as opposed to a possible transitory arrest in the evolution of an observable (panel (e) in Fig.~\ref{fig:Rg_G3_explained}).

In Fig.~\ref{fig:Rg_G3} we apply this analysis more widely to our data for melts of tightly double-folded ring polymers with different values of $N_{\rm ring}$ and $\beta \treetilde\mu_3$.
Consider first columns (a) and (c) showing data equivalent to Fig.~\ref{fig:Rg_G3_explained}(e) for ring and tree simulations respectively, where reversed time axes facilitate the comparison of the attained equilibrium values for the gyration radius. 
While the Amoeba algorithm reaches equilibrium even for the largest rings with $N_{\rm ring}=32768$, this is only the case for $N_{\rm ring}\leq 512$ when following the local ring dynamics. 
Note that after the first MC sweeps (with the earliest data points shown for both algorithms), the gyration radius is already of the order of the equilibrium value for the Amoeba tree sampling and essentially unchanged relative to the starting state if one follows a local dynamics. 

\subsection{Equivalence of the generated ensembles}\label{sec:Validation}
As a first test of the equivalence of the ensembles generated by the algorithms, consider column (b) of Fig.~\ref{fig:Rg_G3}, which shows ring dynamics data, but this time for an ensemble of starting states prepared from equilibrated Amoeba simulations in the tree representation.
Clearly there is no drift whatsoever over the course of these simulations, showing that the present framework allows us to probe the equilibrium dynamics for systems which are much too large to be equilibrated dynamically. 

In Fig.~\ref{fig:Diagonal_plots} we show the correlation between equilibrium values for:
(i) the mean number of branch points ($\langle N_3 \rangle$),
(ii) the mean tree size ($\langle N_{\rm tree} \rangle$)
and
(iii) the mean-square gyration radius ($\langle R_g^2 \rangle$),
for rings of different input parameters ($N_{\rm ring}$, $\beta\treetilde\mu_3$) and different system type (ideal rings, dilute self-avoiding rings and dense melt of rings), obtained using ring dynamics ($x$-axis) and Amoeba sampling ($y$-axis).
The perfect superposition of all datapoints onto the ``$y=x$'' (dashed) line provides further evidence that the two algorithms sample from the same ensembles.
As final validation, the full distribution functions of the corresponding quantities are also in excellent agreement with each other (see insets in the panels of Fig.~\ref{fig:Diagonal_plots}, where only data for melt of rings with $N_{\rm ring}=512$ and $\beta \tilde{\mu}_3=0$ were explicitly shown for clarity). 

\subsection{Efficiency} \label{sec:Efficiency}
For a quantitative comparison of the performances of the two algorithms we estimate the equilibration time $\tau_{\rm eq}$ through the relation:
\begin{equation}\label{eq:equilibration_time_estimator}
\tau_{\rm eq} = \frac{ t_{\rm run}} { g_3(t_{\rm run}) \, / \, \langle R_g^2(t_{\rm run}) \rangle } \, ,
\end{equation}
where $t_{\rm run}$ is the total CPU run time after pruning the corresponding trajectory of the non-equilibrated part ({\it i.e.}, the white areas in panels (a) to (e) in Fig.~\ref{fig:Rg_G3_explained}).
Eq.~\eqref{eq:equilibration_time_estimator}  translates the observation that equilibration occurs when our rings diffuse over a distance comparable to their equilibrium size, 
$g_3(\tau_{\rm eq}) = \lim_{t \to \infty} \langle R_g^2(t) \rangle$, and the fact that $g_3(t)$ behaves diffusively beyond this time.

\begin{figure}
\includegraphics[width=0.48\textwidth]{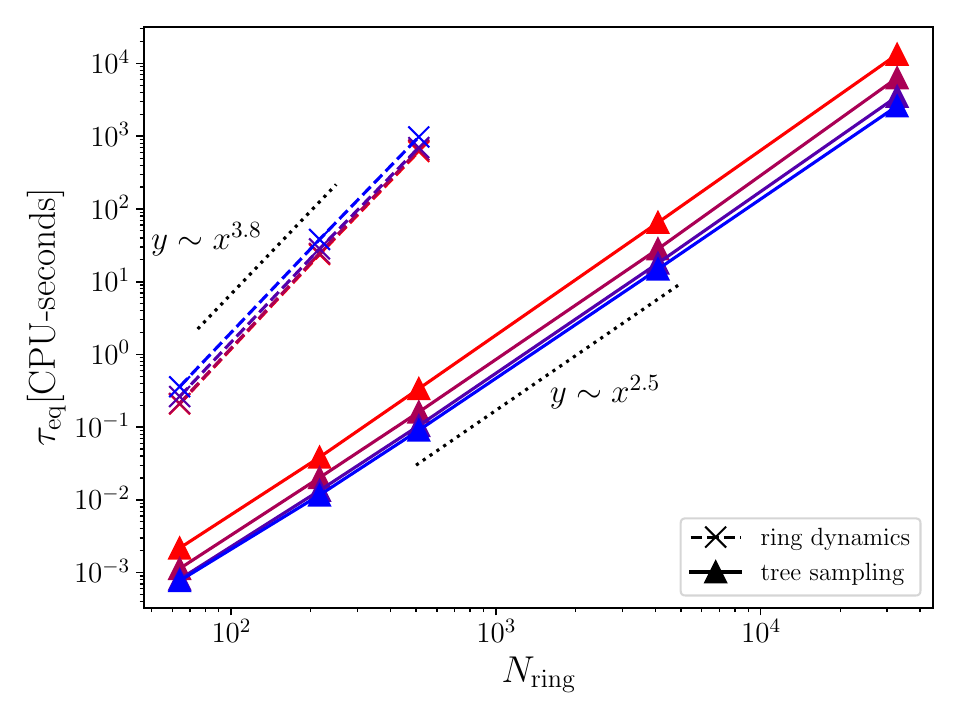}
\caption{
Ring equilibration time $\tau_{\rm eq}$ (in CPU-seconds) for melt simulations using ring dynamics (Sec.~\ref{sec:ElhamsAlgorithm}) and Amoeba tree sampling (Sec.~\ref{sec:SamplingRings}) as a function of ring size $N_{\rm ring}$.
The dotted lines guiding the eye report the observed power law behaviours: $\tau_{\rm eq} \sim {\mathcal O}(N_{\rm ring}^{3.8})$ for dynamic simulations, while $\tau_{\rm eq} \sim {\mathcal O}(N_{\rm ring}^{2.5})$ for Amoeba sampling.
Colorcode is as in Fig.~\ref{fig:Rg_G3}.
}
\label{fig:equilibration_times}
\end{figure}

Results for $\tau_{\rm eq}$ as a function of ring size $N_{\rm ring}$ for melt simulations using the two methods are shown in Fig.~\ref{fig:equilibration_times}.
Apart from a $\treetilde\mu_3$-dependent pre-factor, the equilibration time in the semi-grand canonical Amoeba algorithm scales roughly as $N_{\rm ring}^{2.5}$ (in CPU seconds per polymer), whereas the dynamic MC algorithm is ${\mathcal O}(N_{\rm ring}^{3.8})$, {\it i.e.} at least ${\mathcal O}(N_{\rm ring})$-times slower.

\section{Conclusion and Outlook}\label{sec:Conclusion}
In this work we have established the exact quantitative link between the ensemble of tightly double-folding rings and the ensemble of the underlying random trees (Fig.~\ref{fig:ExampleRing}). 
While based on the counting of microstates for non-interacting systems~\cite{vanderHoek2025,vanderHoekWrapping2026}, the exact correspondence continues to hold in the presence of interactions as long as they are consistently defined across ensembles.
For practical purposes, we have advocated the choice of a one-to-one correspondence between ring and tree conformations allowing to freely switch between the two representations either (i) by eliminating the information on repton locations and the order in which rings wrap the underlying trees or (ii) by re-generating this information via random sampling.  Eqs.~\eqref{eq:w_tree} and~\eqref{eq:muring_mutree} summarise the statistical weights to be used in the tree ensemble to control the number of tree and branch nodes.

In Ref.~\cite{Ghobadpour2025} we have introduced a Monte Carlo algorithm for studying the dynamics of the ring model, where displacements of ring monomers are {\it local} both in space and on the tree. 
Here we have developed an extension of the semi-kinetic Amoeba Monte Carlo scheme~\cite{Amoebapaper2024} for sampling the corresponding tree ensemble, which is based on {\it non-local} mass transport via the insertion and removal of tree nodes at random positions.
While the computational cost for the former algorithm scales like ${\mathcal O}(N_{\rm ring}^{3.8})$, the latter is, at least, a ${\mathcal O}(N_{\rm ring})$-factor faster (see Fig.~\ref{fig:equilibration_times}).
In practice, we can now equilibrate melts of interacting rings with $N_{\rm ring} = 32768$ when previously we had difficulties achieving equilibration for $N_{\rm ring} = 512$. 
Furthermore, we can now easily sample ideal, non-interacting rings of even larger sizes as a useful reference case, since the Pr\"ufer sampling~\cite{vanderHoek2025} of corresponding trees scales as ${\mathcal O}(N_{\rm tree})$ and takes only of the order of one CPU second for $N_\mathrm{tree}=10^6$.

While the present paper is of methodological character, we believe that our integrated approach to the large-scale structure and dynamics of double-folded chain molecules will be relevant in a wide range of genomic and soft-matter systems discussed in the Introduction. Not only are the present results crucial for studying static properties for hitherto inaccessible ring sizes, they also allow us to study at least the short-term dynamics of large rings by initiating simulations from properly (and inexpensively) equilibrated starting states generated on the tree level.

\section*{Supplementary Material}
The supplementary material contains:
a section describing in full details the semi-grand canonical Amoeba algorithm introduced and used in this work;
four tables with details on the parameters and the run times of the systems investigated by ring dynamics and Amoeba tree sampling.

\section*{Acknowledgements}
PHWvdH acknowledges financial support from PNRR\_M4C2I4.1.\_DM351 funded by NextGenerationEU. 
AR acknowledges financial support from PNRR Grant CN\_00000013\_CN-HPC, M4C2I1.4, spoke 7, funded by Next Generation EU.
PHWvdH and AR acknowledge the Ulysses cluster at SISSA (Trieste, Italy), where simulations were performed.
EG and RE gratefully acknowledge the support of the Centre Blaise Pascal's IT test platform at ENS de Lyon (Lyon, France) for facilities.
The platform operates the SIDUS solution~\cite{Quemener2013} developed by Emmanuel Quemener.
All authors acknowledge Ivan Junier and Gabin Laurent for useful discussions.

\section*{Author declarations}
\subsection*{Conflict of interest}
The authors have no conflicts to disclose.

\section*{Data availability}
The data that support the findings of this study are available from the corresponding author upon reasonable request.


\bibliography{../biblio.bib}

\clearpage

\resumetoc
\widetext
\clearpage
\begin{center}
\textbf{\Large Supplemental Material \\
\vspace*{1.5mm}
Coherent modeling of double-folded ring polymers and their underlying random tree structure} \\
\vspace*{5mm}
Pieter H. W. van der Hoek, Angelo Rosa, Elham Ghobadpour, Ralf Everaers
\vspace*{10mm}
\end{center}

\setcounter{equation}{0}
\setcounter{figure}{0}
\setcounter{table}{0}
\setcounter{page}{1}
\setcounter{section}{0}
\setcounter{page}{1}
\makeatletter
\renewcommand{\theequation}{S\arabic{equation}}
\renewcommand{\thefigure}{S\arabic{figure}}
\renewcommand{\thetable}{S\arabic{table}}
\renewcommand{\thesection}{S\arabic{section}}

\tableofcontents



\section{A semi-grand canonical Amoeba algorithm for lattice trees}\label{sec:SemiGCAmoeba}

\subsection{Amoeba for fluctuating tree sizes}\label{sec:GCAmoeba}

\subsubsection{Metropolis-Hastings sampling}
Our objective is to sample an ensemble of embedded trees with a statistical weight
\begin{equation}
w(| s \rangle) = w_{\rm tree}(N_{\rm tree}(| s \rangle), N_{\rm ring}) \, e^{\beta \mu_3 N_3(| s \rangle)} w_{EV}(| s \rangle) 
\end{equation}
using standard Metropolis-Hastings~\cite{Metropolis1953,Hastings} importance sampling.
Given statistical weights $\omega_\alpha$ for choosing among the possible trial moves, the standard detailed balance condition for the equilibrium flux between an initial and a final state, $| i \rangle \Longleftrightarrow | f \rangle$, reads
\begin{equation}\label{eq:standard detailed balance}
w(| i \rangle) \ \alpha(| i \rangle \rightarrow | f \rangle) \ {\rm acc}(| i \rangle \rightarrow | f \rangle) 
	= w(| f \rangle) \ \alpha(| f \rangle \rightarrow | i \rangle) \ {\rm acc}(| f \rangle \rightarrow | i \rangle) 
\end{equation}
with the probability
\begin{equation}
\alpha(| i \rangle \rightarrow | f \rangle) = \frac{\omega_\alpha(|i\rangle \rightarrow |f\rangle)}{z_\alpha\left(|i\rangle\right)}
\end{equation}
for choosing a particular trial move following from the statistical weights of the moves by normalisation with the partition sum of the statistical weights of all possible trial moves in the initial state $| i \rangle$,
\begin{equation}
z_\alpha(| i \rangle) = \sum_{| f \rangle} \omega_\alpha(|i\rangle \rightarrow |f\rangle) \ .
\end{equation}
Reordering yields a condition for the ratio of the acceptance probabilities of the forward and backward move
\begin{equation}\label{eq:ratio of acceptance probabilities}
\frac{{\rm acc}(| i \rangle \rightarrow | f \rangle)}
       { {\rm acc}(| f \rangle \rightarrow | i \rangle)}
=
\frac{w(| f \rangle) \, \, \omega_\alpha(| f \rangle \rightarrow | i \rangle) \, \, z_\alpha(| i \rangle)}
       {w( | i \rangle) \, \, \omega_\alpha(| i \rangle \rightarrow | f \rangle) \, \, z_\alpha(| f \rangle)}
\end{equation}
which is fulfilled by the Metropolis~\cite{Metropolis1953} rates
\begin{equation}\label{acc1}
{\rm acc}_{| i \rangle \rightarrow | f \rangle} = \min \left\{ 1, \, \, \, \frac{w(| f \rangle) \, \, \omega_\alpha(|f\rangle \rightarrow |i\rangle) \, \, z_\alpha(| i \rangle)}
                                                                                                            {w( | i \rangle) \, \, \omega_\alpha(|i\rangle \rightarrow |f\rangle) \, \, z_\alpha(| f \rangle)}\right \} \ .
\end{equation}
When the ratio from the detailed balance condition appearing in Eq.~\eqref{acc1} is of order one, most trial moves are accepted.
However, if the ratio becomes $\ll 1$ for a class of trial moves essential for ergodicity, then the simulation becomes accordingly slower. 

\subsubsection{Amoeba trial moves for fluctuating tree sizes}
The Amoeba algorithm~\cite{SeitzKlein1981,Rosa2016a,Rosa2016b,Amoebapaper2024} proceeds by cutting and pasting ``$f=1$"-functional leaves from and to an embedded tree. The present semi-grand canonical variant is based on two pairs of mutually annihilating move types :
\begin{itemize}
\item \begin{description}
		\item[C2] Cutting a leaf from a ``$f=2$"-functional node, such that $N_2 \rightarrow  N_2 -1$ and  $N_{\rm tree} \rightarrow  N_{\rm tree} -1$.
		\item[P1] Pasting a leaf to one of the $c$ neighbor sites of a ``$f=1$"-functional node, such that $N_2 \rightarrow  N_2 +1$  and  $N_{\rm tree} \rightarrow  N_{\rm tree} +1$.
	\end{description}
\item \begin{description}
		\item[C3] Cutting a leaf from a ``$f=3$"-functional node, such that $N_3 \rightarrow  N_3 -1$, $N_2 \rightarrow  N_2 +1$, $N_1 \rightarrow  N_1 -1$  and  $N_{\rm tree} \rightarrow  N_{\rm tree} -1$.
		\item[P2] Pasting a leaf to one of the $c$ neighbor sites of a ``$f=2$"-functional node, such that $N_3 \rightarrow  N_3 +1$, $N_2 \rightarrow  N_2 -1$, $N_1 \rightarrow  N_1 +1$ and  $N_{\rm tree} \rightarrow  N_{\rm tree} +1$.
	\end{description}
\end{itemize}
In the following we will always choose cutting moves with a statistical weight of $ \omega_{C2} =  \omega_{C3}\equiv 1$ so that
\begin{equation}
z_\alpha(|s\rangle) = 
N_{1}\left(|s\rangle\right) + 
c  \left( N_{1}\left(|s\rangle\right) \omega_{P1} + N_{2}\left(|s\rangle\right) \omega_{P2} \right)
\end{equation}
In particular, the ratios
\begin{equation}
\label{eq:ratio of acceptance probabilities P1C2}
A_{P1/C2} = 
\frac{ {\rm acc}_{P1} }
       { {\rm acc}_{C2} }
=
\frac{w_{\rm tree}(N_{\rm tree}(| i \rangle)+1, N_{\rm ring})}
       {w_{\rm tree}(N_{\rm tree}(| i \rangle)    , N_{\rm ring})} \ 
\frac{w_\mathrm{EV}(| f \rangle)}{w_\mathrm{EV}(| i \rangle)}\ 
\frac{1}{\omega_{P1}} \ 
\frac{z_\alpha\left( N_{1}(|i\rangle),  N_{2}(|i\rangle) \right)}
       {z_\alpha\left( N_{1}(|i\rangle),  N_{2}(|i\rangle)+1 \right)} 
\end{equation}
and
\begin{equation}
\label{eq:ratio of acceptance probabilities P2C3}
A_{P2/C3} = 
\frac{w_{\rm tree}(N_{\rm tree}(| i \rangle)+1, N_{\rm ring})}
       {w_{\rm tree}(N_{\rm tree}(| i \rangle)    , N_{\rm ring})} \ e^{\beta\mu_3}\ 
\frac{w_\mathrm{EV}(| f \rangle)}{w_\mathrm{EV}(| i \rangle)}\ 
\frac{1}{\omega_{P2}} \ 
\frac{z_\alpha\left( N_{1}(|i\rangle),  N_{2}(|i\rangle) \right)}
       {z_\alpha\left( N_{1}(|i\rangle)+1,  N_{2}(|i\rangle)-1 \right)} \ 
\end{equation}
determine the Metropolis acceptance probabilities Eq.~\eqref{acc1} of the four moves through Eq.~\eqref{eq:ratio of acceptance probabilities}
\begin{eqnarray}
{\mathrm acc}_{P1} 
&=& \min \left\{ 1, A_{P1/C2}  \right\} \\
{\mathrm acc}_{C2} 
&=& \min \left\{ 1, 
			\frac1{A_{P1/C2}} 
                \right\} \\
{\mathrm acc}_{P2} 
&=& \min \left\{ 1, 
			A_{P2/C3} 
                \right\} \\
{\mathrm acc}_{C3} 
&=& \min \left\{ 1, 
			\frac{ 1 }
			       { A_{P2/C3}} 
                \right\} 
         \ .
\end{eqnarray}
with the term
\begin{eqnarray}
	\frac{w_{\rm tree}(N_{\rm tree}(| i \rangle)+1, N_{\rm ring})}
	       {w_{\rm tree}(N_{\rm tree}(| i \rangle)    , N_{\rm ring})} 
&=&
	\frac {(2N_{\rm tree}(| i \rangle)-3)! \, (N_{\rm ring}-2N_{\rm tree}(| i \rangle) +2 )!} {(2N_{\rm tree}(| i \rangle)-1)! \, (N_{\rm ring}-2N_{\rm tree}(| i \rangle) )!} \nonumber\\
\label{eq:ControlTreeWeight}
&=&
	\frac {(N_{\rm ring}-2N_{\rm tree}(| i \rangle) +2 )\,(N_{\rm ring}-2N_{\rm tree}(| i \rangle) +1 )} {(2N_{\rm tree}(| i \rangle)-1)\,(2N_{\rm tree}(| i \rangle)-2)}  
\end{eqnarray}
controlling the tree size for $N_{\rm tree}(| i \rangle) \geq2$.

\subsubsection{The limit of large rings and highly branching trees}
To gain some insight into suitable choices for the weights $\omega_{P1}$ and $\omega_{P2}$, it is useful to consider strongly branching double-folded rings and trees with $N_{\rm ring} > N_{\rm tree} > N_{1}=N_3+2,N_2 > 1$ in equilibrium, where 
\begin{eqnarray}
N_{\rm tree} & \approx & \Lambda N_{\rm ring}\\
N_1 \approx N_3 &\approx& \lambda N_{\rm tree} \\
N_2 &\approx& (1-2\lambda) N_{\rm tree} 
\end{eqnarray}
with $0< \lambda < 1/2$ (because there have to be as many branch nodes as tips, their respective fractions cannot exceed $1/2$) and $0< \Lambda < 1/2$ (with the upper limit corresponding to a minimal ring with no reptons wrapping the tree). 
With  
\begin{eqnarray}
\frac{z_\alpha\left( N_{1}(|i\rangle),  N_{2}(|i\rangle) \right)}
       {z_\alpha\left( N_{1}(|i\rangle),  N_{2}(|i\rangle)+1 \right)} 
&=&
	\frac{\left( 1+ c \, \omega_{P1}  \right)N_{1}(|i\rangle) + c \, \omega_{P2} N_{2}(|i\rangle) }
	       {\left( 1+ c \, \omega_{P1} \right) N_{1}(|i\rangle) + c \, \omega_{P2}(N_{2}(|i\rangle)+1) } \\
&\approx&
	\frac{\left( 1+ c \, \omega_{P1}  \right) \lambda + c \, \omega_{P2} (1-2\lambda) }
	       {\left( 1+ c \,  \omega_{P1} \right) \lambda + c \, \omega_{P2}(1-2\lambda + \frac1{N_{\rm tree}} ) } \\
&=&	    {\cal O}(1)   \\
\frac{z_\alpha\left( N_{1}(|i\rangle),  N_{2}(|i\rangle) \right)}
       {z_\alpha\left( N_{1}(|i\rangle)+1,  N_{2}(|i\rangle)-1 \right)} 
&=&
	\frac{\left( 1+ c \, \omega_{P1} \right) N_{1}(|i\rangle) + c \, \omega_{P2} N_{2}(|i\rangle) }
	       {\left( 1+ c \, \omega_{P1 }\right)  (N_{1}(|i\rangle)+1) + c \, \omega_{P2}(N_{2}(|i\rangle)-1) } \\
&\approx&
	\frac{\left( 1+ c \, \omega_{P1} \right)  \lambda + c \, \omega_{P2} (1-2\lambda) }
	       {\left( 1+ c \, \omega_{P1} \right) (\lambda + \frac1{N_{\rm tree}} )+ c \, \omega_{P2}(1-2\lambda - \frac1{N_{\rm tree}} ) } \\
&=&	    {\cal O}(1) 
\end{eqnarray}
we are going to neglect the ratio $\frac{z_\alpha(| i \rangle)}{z_\alpha(| f \rangle)}$ from the subsequent considerations. 
Furthermore,
\begin{eqnarray}
	\frac{w_{\rm tree}(N_{\rm tree}(| i \rangle)+1, N_{\rm ring})}
	       {w_{\rm tree}(N_{\rm tree}(| i \rangle)    , N_{\rm ring})} 
&\approx&
	\frac {(1-2\Lambda +\frac2{N_{\rm ring}} )\,(1-2\Lambda +\frac1{N_{\rm ring}} )} {(2\Lambda-\frac1{N_{\rm ring}})\,(2\Lambda-\frac1{N_{\rm ring}})}  \\
&\approx&
	\frac {(1-2\Lambda )^2} {(2\Lambda)^2}  \\
&\equiv&
	e^{\beta\mu_\mathrm{node}}
\end{eqnarray}
reduces to an approximately constant ``activity'' in equilibrium that controls the tree size in the same way as the factor $e^{\beta\mu_3}$ controls the number of branch nodes.

The node chemical potential is equal to zero for $\Lambda=1/4$ and varies linearly with $\Lambda$ around this value: $\beta\mu_\mathrm{node} = -16 \left( \Lambda - \frac14 \right)$ for $0\ll \Lambda \ll \frac12$.
The behavior in the two limiting case of $\Lambda\rightarrow0$ and $\Lambda\rightarrow1/2$ is easy to understand.
In the latter case, when the ensemble admits only a small number of reptons, it becomes more and more difficult to introduce tree nodes: $\lim_{\Lambda\rightarrow 1/2} \beta\mu_\mathrm{node} = -\infty$. In the opposite limit, when there is a large number of reptons for rings compressed onto a very small trees, node creation is strongly favoured: $\lim_{\Lambda\rightarrow 0} \beta\mu_\mathrm{node} = +\infty$.
In particular, the divergence of $\beta\mu_\mathrm{node}$ in the two limits prevents the removal of the last remaining tree node containing all ring monomers, or the introduction of an additional tree node when the ring is stretched to its maximal extension. 
The term thus accounts for a behavior that occurs naturally when reptons are simulated explicitly: when there are only few reptons along the entire ring, then they are unlikely to be found at a random site to which we are trying to paste a leaf (the dynamic algorithm avoids this problem as the degrees of freedom are not tree nodes but ring monomers).
And when there are many reptons, it is unlikely that there is not a single one present on a to-be-deleted tree node.

\subsubsection{Naive choice of the trial moves}
Consider first a naive choice of the trial moves with $ \omega_{C2} =  \omega_{C3}  \equiv 1$ and 
\begin{eqnarray}
\omega_{P1} &=& 1\\
\omega_{P2} &=& 1 \ ,
\end{eqnarray}
so that
\begin{equation}
z(|s\rangle) = \left( 1+ c  \right) N_{1}(|s\rangle) + c  N_{2}(|s\rangle) \ .
\end{equation}
In this case, large node chemical potentials, $\left|\beta\mu_\mathrm{node}\right| \gg 0$, can strongly reduce the acceptance rates for pasting a node to a leaf 
\begin{equation}\label{accP1}
{\mathrm acc}_{P1} \approx \min \left\{ 1, 
				e^{\beta\mu_\mathrm{node}}
				\frac{w_\mathrm{EV}(| f \rangle)}{w_\mathrm{EV}(| i \rangle)}
                          \right\}
\ .
\end{equation}
or for cutting a leaf from a $f=2$ functional node
\begin{equation}\label{accC2}
{\mathrm acc}_{C2} \approx \min \left\{ 1, 
				e^{-\beta\mu_\mathrm{node}} 
				\frac{w_\mathrm{EV}(| f \rangle)}{w_\mathrm{EV}(| i \rangle)}
                          \right\}
\ .
\end{equation}
Independently, large branching chemical potentials, $\left|\beta\mu_3\right| \gg 0$, can induce high rejection rates for attempts to create a new branch point through the pasting of a leaf to a ``$f=2$''-functional node, 
\begin{equation}\label{accP2}
{\mathrm acc}_{P2} \approx \min \left\{ 1, 
				e^{\beta\mu_\mathrm{node}}\ e^{\beta\mu_3}\ 
				\frac{w_\mathrm{EV}(| f \rangle)}{w_\mathrm{EV}(| i \rangle)}
                          \right\}
\ ,
\end{equation}
or attempts to create a ``$f=2$''-functional node by removing a leaf from a branch point,
\begin{equation}\label{accC3}
{\mathrm acc}_{C3} \approx \min \left\{ 1, 
				e^{-\beta\mu_\mathrm{node}}\ e^{-\beta\mu_3}\ 
				\frac{w_\mathrm{EV}(| f \rangle)}{w_\mathrm{EV}(| i \rangle)}
                          \right\}
\ .
\end{equation}
One can deal with these problems along the lines of our ``semi-kinetic'' algorithm from Ref.~\cite{Amoebapaper2024} by a suitable reduction of the attempt frequency of often-rejected trial moves.

\subsubsection{Including the control of the branching activity into the attempt frequencies for the trial moves}
Consider first the choice $ \omega_{C2} =  \omega_{C3} \equiv 1$ and 
\begin{eqnarray}
\omega_{P1} &=& 1\\
\omega_{P2} &=& e^{\beta\mu_3} \ ,
\end{eqnarray}
so that
\begin{equation}
z(|s\rangle) = \left( 1+ c  \right) N_{1}(|s\rangle) + c\,  e^{\beta\mu_3} N_{2}(|s\rangle) \ .
\end{equation}
The weight $\omega_{P2} $ cancels the factor $e^{\beta\mu_3}$ from the ratio of the statistical weights for trees with $N_3+1$ and $N_3$ branch points in Eq.~\eqref{eq:ratio of acceptance probabilities P2C3}.
With Eqs.~\eqref{eq:ratio of acceptance probabilities P1C2} and~\eqref{eq:ratio of acceptance probabilities P2C3} being equal, the acceptance rates for the two types of pasting and the two types of cutting moves become identical: 
\begin{eqnarray}
\label{accP1accP2 semi kinetic-1}
{\mathrm acc}_{P1} = {\mathrm acc}_{P2}
&\approx& \min \left\{ 1, 
				e^{\beta\mu_\mathrm{node}}
				\frac{w_\mathrm{EV}(| f \rangle)}{w_\mathrm{EV}(| i \rangle)}
                          \right\}\\
\label{accC2accC3 semi kinetic-1}
{\mathrm acc}_{C2} = {\mathrm acc}_{C3}
&\approx& \min \left\{ 1, 
				e^{-\beta\mu_\mathrm{node}} 
				\frac{w_\mathrm{EV}(| f \rangle)}{w_\mathrm{EV}(| i \rangle)}
                          \right\}
\ .
\end{eqnarray}
%

\subsubsection{Including the control of the number of tree nodes into the attempt frequencies for the trial moves}\label{sec:efficient_choice}
One can push this reasoning one step further with the choice $ \omega_{C2} =  \omega_{C3}  \equiv 1$ and 
\begin{eqnarray}
\omega_{P1} 
&=& 	\frac{w_{\rm tree}(N_{\rm tree}(| i \rangle)+1   , N_{\rm ring})}
	        {w_{\rm tree}(N_{\rm tree}(| i \rangle), N_{\rm ring})}
\\
\omega_{P2} 
&=& 	\frac{w_{\rm tree}(N_{\rm tree}(| i \rangle)+1    , N_{\rm ring})}
	        {w_{\rm tree}(N_{\rm tree}(| i \rangle), N_{\rm ring})}
e^{\beta\mu_3} \ ,
\end{eqnarray}
with the ratio of the statistical weights for trees with $N_\mathrm{tree}+1$ and $N_\mathrm{tree}$ tree nodes written out in Eq.~\eqref{eq:ControlTreeWeight}.
In this case
\begin{equation}
z(|s\rangle) = N_{1}(|s\rangle) + c  \frac{w_{\rm tree}(N_{\rm tree}(| i \rangle)+1    , N_{\rm ring})}
	                                                           {w_{\rm tree}(N_{\rm tree}(| i \rangle), N_{\rm ring})}
	                                                    \left( N_{1} +   e^{\beta\mu_3} N_{2}(|s\rangle) \right) 
\end{equation}
In addition to the factor controlling the number of branch points, the weights $\omega_{P1}$ and $\omega_{P2}$ also cancel the factor controlling the number of tree nodes in Eqs.~(\ref{eq:ratio of acceptance probabilities P1C2}) and (\ref{eq:ratio of acceptance probabilities P2C3}). As a consequence, all acceptance rates are now solely controlled by the excluded volume part of the Hamiltonian: 
\begin{eqnarray}
\label{accP1accP2 semi kinetic-2}
{\mathrm acc}_{P1} = {\mathrm acc}_{P2}
&\approx& \min \left\{ 1, 
				\frac{w_\mathrm{EV}(| f \rangle)}{w_\mathrm{EV}(| i \rangle)}
                          \right\}\\
\label{accC2accC3 semi kinetic-2}
{\mathrm acc}_{C2} = {\mathrm acc}_{C3}
&\approx& \min \left\{ 1, 
				\frac{w_\mathrm{EV}(| f \rangle)}{w_\mathrm{EV}(| i \rangle)}
                          \right\}
\ .
\end{eqnarray}
%

\subsubsection{Choosing cut and paste moves with equal probability}\label{sec:chosen_convention}
As an alternative, consider the choice $ \omega_{C2} = \omega_{C3} \equiv 1$ and 
\begin{eqnarray} 
\omega_{P1} \label{eq:weight_chosen1}
&=& 	\frac1c
\\
\omega_{P2} \label{eq:weight_chosen2}
&=& 	\frac1c
e^{\beta\mu_3} \ ,
\end{eqnarray}
so that
\begin{equation}\label{eq:move_z_used}
z(|s\rangle) = 2 N_{1}(|s\rangle) + e^{\beta\mu_3} N_{2}(|s\rangle)  \ .
\end{equation}
Suppressing the statistical weight of paste moves by a factor of $c$ implies that cut and paste moves are attempted with equal frequencies.
In this case
\begin{eqnarray}
\label{accP1accP2 semi kinetic-3}
{\mathrm acc}_{P1} = {\mathrm acc}_{P2}
&\approx& \min \left\{ 1, 
				e^{\beta\mu_\mathrm{node}}
				\,c\,
				\frac{w_\mathrm{EV}(| f \rangle)}{w_\mathrm{EV}(| i \rangle)}
                          \right\}\\
\label{accC2accC3 semi kinetic-3}
{\mathrm acc}_{C2} = {\mathrm acc}_{C3}
&\approx& \min \left\{ 1, 
				e^{-\beta\mu_\mathrm{node}} 
				\,\frac1c\,
				\frac{w_\mathrm{EV}(| f \rangle)}{w_\mathrm{EV}(| i \rangle)}
                          \right\}
\ .
\end{eqnarray}
This is the implementation used in the present work. 
For all simulated ensembles we observed equilibrium values in the range $0.29 \lesssim \Lambda \lesssim 0.43$. 
For $c=12$, this implies that the factor $e^{-\beta\mu_\mathrm{node}}/c$ remains of order unity, so that the resulting acceptance-rate imbalance affects the equilibration times in Fig.~\ref{fig:equilibration_times} of main text only by a prefactor of comparable magnitude. 
The choice of weights discussed in Sec.~\ref{sec:efficient_choice} nevertheless provides a natural route to a more efficient implementation, especially for ensembles in which $\Lambda$ approaches the limiting regimes.

\subsubsection{Including single-node states}\label{sec:Singlenode}
Although unlikely, the limiting state $| N_{\rm tree}=1\rangle$ is dynamically accessible when explicitly simulating ring monomers.
However, in this state the single tree node has nominal functionality $f=0$, therefore it does not fall within the discussion presented so far.
To fix this issue, here we treat conventionally the single node as if it were a ``$f=1$''-node, implying (see Eq.~\eqref{eq:move_z_used})
\begin{equation}\label{eq:Z_small}
z\left(| N_{\rm tree}=1\rangle\right) = 2 \, .
\end{equation}
Then, using the weights introduced in Sec.~\ref{sec:chosen_convention}, the following acceptance probabilities follow:
\begin{equation}\label{eq:acc4+5+6}
{\rm acc}_{| i \rangle \rightarrow | f \rangle} = \left\{
\begin{array}{cc}
\min \left\{ 1, \, \, \, \frac2{N_{\rm ring}\,(N_{\rm ring}-1 )} \, \frac1c \right  \} , & \mbox{for leaf-cutting when } N_{\rm tree}(|i\rangle) = 2 \\
\\
\min \left\{ 1, \, \, \, \frac{N_{\rm ring}\,(N_{\rm ring}-1 )}2 \, c \right  \} , & \mbox{for leaf-pasting when } N_{\rm tree}(|i\rangle) = 1 \\
\\
0 , & \mbox{for leaf-cutting when } N_{\rm tree}(|i\rangle) = 1 
\end{array}
\right.
\end{equation}
which conclude the discussion. 

\subsection{Data structures}\label{sec:DataStructures}
An implementation of the algorithm requires data structures (arrays) with the following information:
\begin{itemize}
\item
Coordinates of all $N_{\rm tree}$ ``active'' nodes.
\item
All nodes with functionality ``$f=1$''.
\item
All nodes with functionality ``$f=2$''.
\item
All nodes with functionality ``$f=3$''.
\item 
All node-labels which are unused ({\it i.e.} the auxiliary list {\it UnusedNodes})
\item
A representation of the connectivity of the $N_{\rm tree}$ ``active'' nodes.
\end{itemize}
In order to optimize performance, the following data-structures can be used additionally:
\begin{itemize}
\item
Functionality of the tree nodes.
\item
Indices of all tree nodes in the ``$f=1$''-, ``$f=2$''- and ``$f=3$''-arrays.
\item
The occupation of every lattice site, to keep track of excluded volume interactions (see Sec.~\ref{sec:IncludingInteractions} of main text).

\end{itemize}
Given a viable starting state ({\it e.g.}, several polymers of size $N_{\rm tree}=1$ on a lattice), the algorithm only requires $N_{\rm ring}$ and $\treetilde\mu_3$ to generate the proper ensemble.

\subsection{Algorithm's implementation}\label{sec:TheAlgo}
For given $N_{\rm ring}$ and $\tilde\mu_3$, we introduce the following algorithm for dense solutions of $P$ lattice trees:
\begin{enumerate}
\item
Pick a polymer $1\leq p \leq P$ at random;
\item
Compute $z(|p\rangle)$ using Eq.~\eqref{eq:move_z_used};
\item
Draw a random number $0 \leq rn <  z(|p\rangle)$, then:
\begin{itemize}
\item[(3.1)]
For $rn < N_1$:
Pick randomly a leaf of the tree and remove it with probability indicated by Eq.~\eqref{acc1} and conventions as in Secs.~\ref{sec:chosen_convention} and~\ref{sec:Singlenode}.
If the move is accepted, add the label of the removed leaf to the auxiliary list {\it UnusedNodes}. 
Update all other datasets accordingly.
\item[(3.2)]
For $N_1 \leq rn < 2N_1$ (respectively, $2N_1 \leq rn <  z(|p\rangle)$):
Place a node at random on one of the $c$ nearest positions of a randomly picked ``$f=1$''-node (resp., ``$f=2$''-node), with probability indicated by Eq.~\eqref{acc1} and conventions as in Secs.~\ref{sec:chosen_convention} and~\ref{sec:Singlenode}. 
If the move is accepted, the label of the newly placed node is randomly picked from {\it UnusedNodes}.
Update all other datasets accordingly.
\end{itemize}
\item
Repeat steps $(3.1)$ and~$(3.2)$ for a total of MC steps $t_{\rm MC} = \tau_{\rm eq}$, where $\tau_{\rm eq}$ corresponds to the equilibration time of the ring as defined in Sec.~\ref{sec:Efficiency} of main text.
\item
At the end of the simulation, determine for each tree the number of zero-length ring monomers $N_{\rm rept}$ from the size of the generated tree (see Eq.~\eqref{eq:Nrep} of main text), and distribute them at random along a ring of size $N_{\rm ring}$.
Wrap the generated tree with the remaining non-zero length bonds following the procedure described in Fig.~2 of our work~\cite{vanderHoekWrapping2026} and insert the zero-length bonds at the sampled locations on the ring. 
\end{enumerate}
%

\clearpage

\section{Simulation input parameters}\label{sec:input}

\subsection{First run: Ideal simulations}

\begin{table*}[h]
\begin{tabular}{|c| c| c| c| c| c|}
\hline
\multicolumn{2}{|c|}{} & \multicolumn{2}{c|}{Ring dynamics} & \multicolumn{2}{c|}{Amoeba tree sampling} \\
\hline
$N_{\mathrm{ring}}$ & $\beta \tilde{\mu}_3$ & Sample size& Run time (MC sweeps) & Sample size& Run time (MC steps per polymer) \\
\hline
64 & $0$ & $200$ & $10^7$ &1200 &$6\times 10^8 $\\
\hline
64 & $-1$ & $200$ & $10^7$ &1200 &$6\times 10^8 $\\
\hline
64 & $-2$ & $200$ & $10^7$ &1200 &$6\times 10^8 $\\
\hline
64 & $-3$ & $200$ & $10^7$ &1200 &$6\times 10^8 $\\
\hline
64 & $-4$ & $200$ & $10^7$ &1200 &$6\times 10^8 $\\
\hline
512 & $0$ & $200$ & $10^8$ &1200 &$6\times 10^8 $\\
\hline
512 & $-2$ & $200$ & $10^8$ &1200 &$6\times 10^8 $\\
\hline
512 & $-4$ & $200$ & $10^8$ &1200 &$6\times 10^8 $\\
\hline
512 & $-6$ & $200$ & $10^8$ &1200 &$6\times 10^8 $\\
\hline
\end{tabular}
\caption{}
\end{table*}

\clearpage
\subsection{First run: Self-avoiding simulations}

\begin{table*}[h]
\begin{tabular}{|c|c|c|c|c|c|c|c|}
\hline
\multicolumn{4}{|c|}{} & \multicolumn{2}{c|}{Ring dynamics} & \multicolumn{2}{c|}{Amoeba tree sampling} \\
\hline
$N_{\mathrm{ring}}$ & $\beta \tilde{\mu}_3$ & $N_{\mathrm{pol}}$ & \#cells (FCC lattice) & Sample size & Run time (MC sweeps) & Sample size & Run time (MC steps per polymer) \\
\hline
64 & $0$ & 1 & $ (20\times 20\times 20)$ & $200$ & $10^7$ &1200 &$6\times 10^8 $\\
\hline
64 & $-1$ & 1 & $ (20\times 20\times 20)$ & $200$ & $10^7$ &1200 &$6\times 10^8 $\\
\hline
64 & $-2$ & 1 & $ (20\times 20\times 20)$ & $200$ & $10^7$ &1200 &$6\times 10^8 $\\
\hline
64 & $-3$ & 1 & $ (20\times 20\times 20)$ & $200$ & $10^7$ &1200 &$6\times 10^8 $\\
\hline
64 & $-4$ & 1 & $ (20\times 20\times 20)$ & $200$ & $10^7$ &1200 &$6\times 10^8 $\\
\hline
512 & $0$ & 1 & $ (40\times 40\times 40)$ & $200$ & $10^8$ &1200 &$6\times 10^8 $\\
\hline
512 & $-2$ & 1 & $ (40\times 40\times 40)$ & $200$ & $10^8$ &1200 &$6\times 10^8 $\\
\hline
512 & $-4$ &1 & $ (40\times 40\times 40)$ & $200$ & $10^8$ &1200 &$6\times 10^8 $\\
\hline
512 & $-6$ &1 & $ (40\times 40\times 40)$ & $200$ & $10^8$ &1200 &$6\times 10^8 $\\
\hline
\end{tabular}
\caption{}
\end{table*}

\clearpage
\subsection{First run: Melt simulations}

\begin{table*}[h]
\begin{tabular}{|c|c|c|c|c|c|c|c|}
\hline
\multicolumn{4}{|c|}{} & \multicolumn{2}{c|}{Ring dynamics} & \multicolumn{2}{c|}{Amoeba tree sampling} \\
\hline
$N_{\mathrm{ring}}$ & $\beta \tilde{\mu}_3$ & $N_{\mathrm{pol}}$ & \#cells (FCC lattice) & $N_{\mathrm{boxes}}$ & Run time (MC sweeps)& $N_{\mathrm{boxes}}$ & Run time (MC steps per polymer) \\
\hline
64 & $0$ & 12 & $ (4\times 4\times 4)$ & $100$ & $10^7$ & 200 & $6\times 10^8 $\\
\hline
64 & $-1$ & 12 & $  (4\times 4\times 4)$ & $100$ & $10^7$ & 200 & $6\times 10^8 $\\
\hline
64 & $-2$ & 12 & $  (4\times 4\times 4)$ & $100$ & $10^7$ & 200 & $6\times 10^8 $\\
\hline
64 & $-3$ & 12 & $  (4\times 4\times 4)$ & $100$ & $10^7$ & 200 & $6\times 10^8 $\\
\hline
64 & $-4$ & 12 & $  (4\times 4\times 4)$ & $100$ & $10^7$ & 200 & $6\times 10^8 $\\
\hline
216 & 0 & 12 & $ (6\times 6\times 6)$ & $100$ & $10^8$ & 200 &$6\times 10^8 $\\
\hline
216 & $-1$ & 12 & $ (6\times 6\times 6)$ & $100$ & $10^8$ & 200 &$6\times 10^8 $\\
\hline
216 & $-2$ & 12 & $ (6\times 6\times 6)$ & $100$ & $10^8$ & 200 &$6\times 10^8 $\\
\hline
216 & $-4$ & 12 & $ (6\times 6\times 6)$ & $100$ & $10^8$ & 200 &$6\times 10^8 $\\
\hline
216 & $-6$ & 12 & $ (6\times 6\times 6)$ & $100$ & $10^8$ & 200 &$6\times 10^8 $\\
\hline
512 & $0$ & 12 & $ (8\times 8\times 8)$ & $100$ & $10^8$ & 200 &$6\times 10^8 $\\
\hline
512 & $-2$ & 12 & $ (8\times 8\times 8)$ & $100$ & $10^8$ & 200 &$6\times 10^8 $\\
\hline
512 & $-4$ &12 & $ (8\times 8\times 8)$ & $100$ & $10^8$ & 200 &$6\times 10^8 $\\
\hline
512 & $-6$ &12 & $ (8\times 8\times 8)$ & $100$ & $10^8$ & 200 &$6\times 10^8 $\\
\hline
\end{tabular}
\caption{}
\end{table*}

\clearpage
\subsection{Second run: Melt simulations}

\begin{table*}[h]
\begin{tabular}{|c|c|c|c|c|c|c|c|}
\hline
\multicolumn{4}{|c|}{} & \multicolumn{2}{c|}{Ring dynamics$^1$} & \multicolumn{2}{c|}{Amoeba tree sampling} \\
\hline
$N_{\mathrm{ring}}$ & $\beta \tilde{\mu}_3$ & $N_{\mathrm{pol}}$ & \#cells (FCC lattice) &  $N_{\mathrm{boxes}}$  & Run time (MC sweeps)& $N_{\mathrm{boxes}}$  & Run time (MC steps per polymer) \\
\hline
64 & $0$ & 12 & $ (4\times 4\times 4)$ &100 & $1\times 10^8 $ &100 & $6\times 10^8 $\\
\hline
64 & $-2$ & 12 & $  (4\times 4\times 4)$ &100 & $1\times 10^8 $&100 & $6\times 10^8 $\\
\hline
64 & $-4$ & 12 & $  (4\times 4\times 4)$ & 100 & $1\times 10^8 $&100 & $6\times 10^8 $\\
\hline
64 & $-6$ & 12 & $  (4\times 4\times 4)$ &100 & $1\times 10^8 $&100 & $6\times 10^8 $\\
\hline
216 & $0$ & 12 & $ (6\times 6\times 6)$ &100 & $1\times 10^8 $ &100 & $6\times 10^8 $\\
\hline
216 & $-2$ & 12 & $ (6\times 6\times 6)$ & 100 & $1\times 10^8 $ &100 & $6\times 10^8 $\\
\hline
216 & $-4$ & 12 & $ (6\times 6\times 6)$ & 100 & $1\times 10^8 $ &100 & $6\times 10^8 $\\
\hline
216 & $-6$ & 12 & $ (6\times 6\times 6)$ &100 & $1\times 10^8 $&100 & $6\times 10^8 $\\
\hline
512 & $0$ & 12 & $ (8\times 8\times 8)$ & 100 & $6.4 \times 10^7 $&100 & $6\times 10^8 $\\
\hline
512 & $-2$ & 12 & $ (8\times 8\times 8)$ & 100 & $6.4\times 10^7 $ &100 & $6\times 10^8 $\\
\hline
512 & $-4$ &12 & $ (8\times 8\times 8)$ & 100 & $6.4\times 10^7 $  &100 & $6\times 10^8 $\\
\hline
512 & $-6$ &12 & $ (8\times 8\times 8)$ & 100 & $6.4\times 10^7 $ &100 & $6\times 10^8 $\\
\hline
4096 & $0$ & 12 & $ (16\times 16\times 16)$ &100 & $8\times 10^6 $ &100 & $1.8\times 10^{10} $\\
\hline
4096& $-2$ & 12 & $ (16\times 16\times 16)$ &100 & $8\times 10^6 $  &100 & $1.8\times 10^{10}  $\\
\hline
4096 & $-4$ &12 & $ (16\times 16\times 16)$  &100 & $8\times 10^6 $  &100 & $1.8\times 10^{10}  $\\
\hline
4096 & $-6$ &12 & $ (16\times 16\times 16)$ & 100 & $8\times 10^6 $  &100 & $1.8\times 10^{10}$ \\
\hline
32768 & $0$ & 12 & $ 32\times 32\times 32)$ & 100 & $5\times 10^5 $&100 & $8.1\times 10^{10} $\\
\hline
32768& $-2$ & 12 & $ (32\times 32\times 32)$ & 100 & $  5\times 10^5$  &100 &$8.1\times 10^{10}  $\\
\hline
32768& $-4$ &12 & $ (32\times 32\times 32)$ & 100 & $5\times 10^5 $ &100 & $8.1\times 10^{10} $\\
\hline
32768 & $-6$ &12 & $ (32\times 32\times 32)$ & 100 & $5\times 10^5$ & 100 & $8.1\times 10^{10} $\\
\hline
\end{tabular}
\caption{
$^1$For ring dynamics simulations, we adopted the same input parameters both when starting from a compact initial state and when starting from an equilibrated state prepared according to the Amoeba tree sampling protocol. 
}
\end{table*}

\end{document}